\documentclass[aps, prb, amsmath, amssymb, english, twocolumn,reprint,floatfix, superscriptaddress]{revtex4-2}

\usepackage[english]{babel}
\usepackage{graphicx}
\usepackage{verbatim}

\usepackage{epsfig}
\usepackage{epstopdf}
\usepackage{amsfonts}
\usepackage{amsthm}
\usepackage{amsmath}
\usepackage{amssymb}
\usepackage{color}
\usepackage[usenames,dvipsnames,svgnames,table]{xcolor}
\usepackage{hyperref} 
\hypersetup{pdfpagemode=FullScreen,colorlinks=true,linkcolor=NavyBlue,citecolor=BrickRed}
\usepackage{makeidx}
\usepackage{pifont}

\usepackage{booktabs}
\usepackage{xr}
\usepackage{color}
\usepackage{grffile}
\usepackage{physics}

\usepackage[ruled]{algorithm2e}
\usepackage[noend]{algpseudocode}
\usepackage{multirow}
\usepackage{type1cm} % used to replace {wasysym}
\usepackage{dcolumn}% Align table columns on decimal point
\usepackage{bm}% bold math

\begin{document}

\title{Electronic structure and correlation of La$_4$Co$_2$NiO$_8$Cl$_2$: a theoretical proposal for a La$_4$Ni$_3$O$_{10}$-like high-temperature superconductor}
% Force line breaks with \\
\author{Si-Yong Jia}\thanks{These authors contributed equally to this work.}\affiliation{School of Physics and Beijing Key Laboratory of Opto-electronic Functional Materials $\&$ Micro-nano Devices, Renmin University of China, Beijing 100872, China}\affiliation{Key Laboratory of Quantum State Construction and Manipulation (Ministry of Education), Renmin University of China, Beijing 100872, China}\affiliation{School of Science, China University of Mining and Technology, Beijing 100083, China}
\author{Jing-Xuan Wang}\thanks{These authors contributed equally to this work.}\affiliation{School of Physics and Beijing Key Laboratory of Opto-electronic Functional Materials $\&$ Micro-nano Devices, Renmin University of China, Beijing 100872, China}\affiliation{Key Laboratory of Quantum State Construction and Manipulation (Ministry of Education), Renmin University of China, Beijing 100872, China}
\author{Jian-Hong She}\affiliation{School of Physics and Beijing Key Laboratory of Opto-electronic Functional Materials $\&$ Micro-nano Devices, Renmin University of China, Beijing 100872, China}\affiliation{Key Laboratory of Quantum State Construction and Manipulation (Ministry of Education), Renmin University of China, Beijing 100872, China}
\author{Rong-Qiang He}\email{rqhe@ruc.edu.cn}\affiliation{School of Physics and Beijing Key Laboratory of Opto-electronic Functional Materials $\&$ Micro-nano Devices, Renmin University of China, Beijing 100872, China}\affiliation{Key Laboratory of Quantum State Construction and Manipulation (Ministry of Education), Renmin University of China, Beijing 100872, China}
\author{Zhong-Yi Lu}\email{zlu@ruc.edu.cn}\affiliation{School of Physics and Beijing Key Laboratory of Opto-electronic Functional Materials $\&$ Micro-nano Devices, Renmin University of China, Beijing 100872, China}\affiliation{Key Laboratory of Quantum State Construction and Manipulation (Ministry of Education), Renmin University of China, Beijing 100872, China}\affiliation{Hefei National Laboratory, Hefei 230088, China}

\date{\today}% It is always \today, but any date may be explicitly specified

\begin{abstract}
Based on the discovery of high-temperature superconductivity in the bilayer nickelate La$_3$Ni$_2$O$_7$, several Co-based La$_3$Ni$_2$O$_7$-like materials were theoretically predicted as possible high-temperature superconductors by electron doping. Motivated by these findings and the subsequent discovery of superconductivity in the trilayer nickelate La$_4$Ni$_3$O$_{10}$ under high pressure, we propose and investigate a Co-based La$_4$Ni$_3$O$_{10}$-like material. With electron doping to the high-pressure trilayer cobaltate La$_4$Co$_3$O$_{10}$, using density functional theory combined with dynamical mean-field theory (DFT+DMFT), we find that the resulting compound La$_4$Co$_2$NiO$_8$Cl$_2$ exhibits a crystal structure and a strongly correlated electronic structure similar to those of La$_4$Ni$_3$O$_{10}$ under high pressure. This suggests that this new compound may host high-temperature superconductivity.

\end{abstract}

% provides a new class of research tools~\cite{Jordan2015science,Carleo2019RMP}
% contrastive learning of visual representations.~\cite{2020SimCLR,wang2022molecular}

%\keywords{Suggested keywords}%Use showkeys class option if keyword display desired
\maketitle

%\tableofcontents
%%%%%%%%%%%%%%%%%%%%%%%%%%%%%%%%%%%%%%%%%%%%%%%%%%%%%%%%%%%%%%%%%%%%%%%%%
%%--------------------------------- Section 1 ----------------------------------%%
%%%%%%%%%%%%%%%%%%%%%%%%%%%%%%%%%%%%%%%%%%%%%%%%%%%%%%%%%%%%%%%%%%%%%%%%%
\section{INTRODUCTION}
The discovery of high-temperature superconductivity in cuprates has motivated extensive exploration of superconducting materials in other 3d transition-metal compounds.~\cite{chu1987superconductivity,schilling1993superconductivity,putilin1993superconductivity,park1995structures,kamihara2008iron,rotter2008superconductivity,johnston2010puzzle}. Among these, nickelate superconductors were first theoretically predicted by analogy with cuprates and subsequently confirmed experimentally~\cite{botana2020similarities,poltavets2009electronic,greenblatt2010bulk}. And chemical substitution has been widely used to tune electronic properties in correlated oxides, including cuprates, iron-based superconductors, and nickelates~\cite{armitage2010progress,sakakibara2020model,graser2009near,kemper2010sensitivity}. Due to the success of this approach, Wang et al.~\cite{wang2025prediction} theoretically predicted several Co-based La$_3$Ni$_2$O$_7$-like materials as possible high-temperature superconductors, based on the recently discovered bilayer Ruddlesden-Popper (RP) phase La$_3$Ni$_2$O$_7$ with a superconductivity transition temperature up to 80 K under high pressure~\cite{sun2023signatures,wang2024pressure,zhou2025investigations,
luo2023bilayer}. Accordingly, the subsequent discovery of superconductivity in the trilayer RP phase La$_4$Ni$_3$O$_{10}$ (LNO-4310) naturally serves as another promising analog for exploring Co-based superconductors~\cite{zhu2024superconductivity,li2024signature}. 

%%%%The emergence of superconductivity in both systems has spurred extensive theoretical investigations into their crystal structures and electronic correlation properties~\cite{shilenko2023correlated,cao2024flat,liu2023s,chen2024evidence}.
For LNO-4310, the crystal structure, electronic structure and correlation are crucial for understanding its superconductivity and have been the subject of extensive studies. The crystal structure undergoes a transition from the monoclinic $P21/a$ to the tetragonal $I4/mmm$ phase in the pressure range of 12.6 to 13.4 GPa~\cite{li2023structural}. Notably, zero resistance is observed above 43.0 GPa, with a maximum $T_c$ of approximately 30 K~\cite{zhu2024superconductivity}. The electronic structure of LNO-4310 is characterized by strong electron correlation and orbital selectivity ~\cite{leonov2024electronic,sakakibara2024theoretical,sakakibara2024possible}. Previous density functional theory plus dynamical mean-field theory (DFT+DMFT) studies have revealed that the correlation in LNO-4310 is both layer-dependent and orbital-dependent~\cite{wang2024non}. Specifically, the outer layers exhibit stronger electronic correlation and non-Fermi liquid behavior with a $T$-linear scattering rate, which is linked to Hund's metal physics and may explain the ``strange metal'' behavior observed in transport experiments~\cite{li2023structural,
li2024signature,zhang2025superconductivity}. In contrast, the inner layer is more hole-doped and displays Fermi-liquid characteristics~\cite{wang2024non}. These findings highlight the importance of layer-resolved electronic structures in understanding the physical properties of LNO-4310. 

\begin{figure}[htb]
  \centering
  \includegraphics[width=8.6cm]{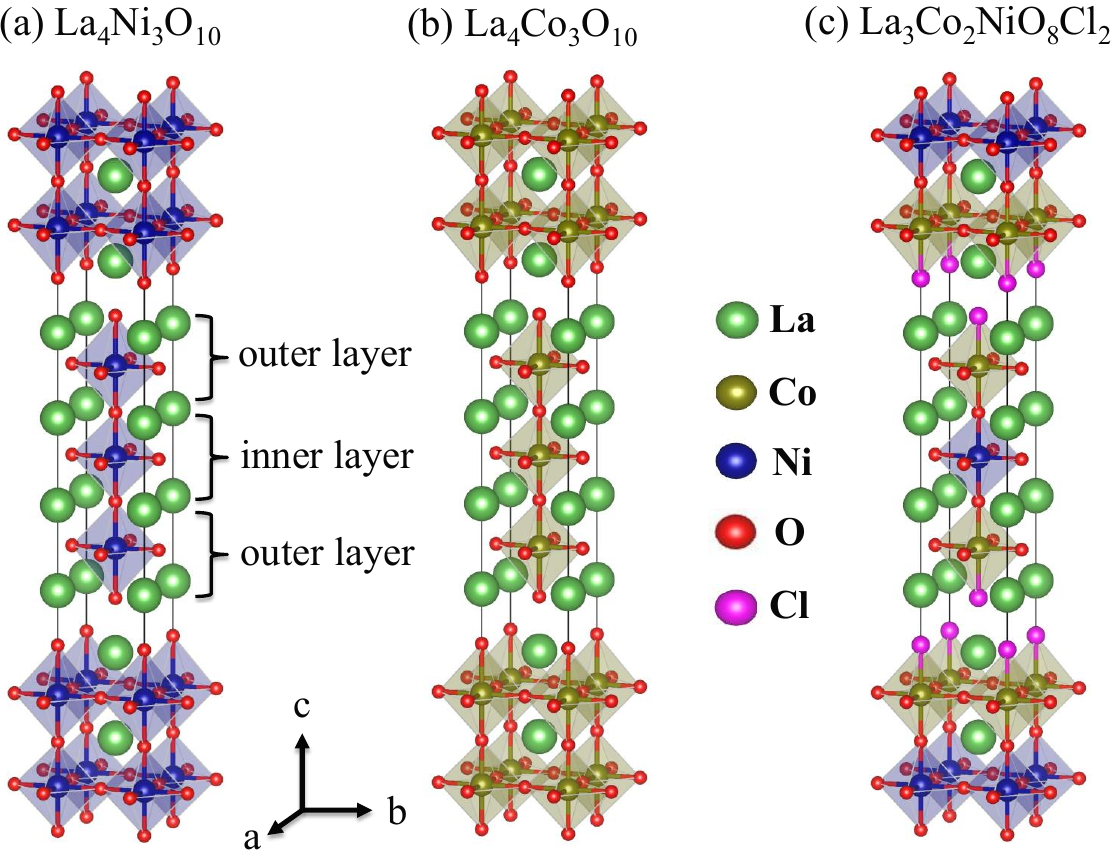}
  \caption{Crystal structure of (a) LNO-4310, (b) LCO-4310 and (c)  LCO-NiCl.}
  \label{fig:structure}
\end{figure}

Chemical substitution has also been shown to be an effective means of tuning electronic structure and correlation~\cite{hu2015predicting}. Therefore, similar to the tetragonal $I4/mmm$ crystal structure of LNO-4310 shown in Fig.~\ref{fig:structure}(a), we apply this strategy to the La$_4$Co$_3$O$_{10}$ (LCO-4310) structure illustrated in Fig.~\ref{fig:structure}(b). The Co atom in LCO-4310 has a valence of +2.67 with a 3d$^{6.33}$ electronic configuration, whereas the Ni atom in LNO-4310 has a 3d$^{7.33}$ configuration. Considering that in LNO-4310 the inner layer is more hole-doped~\cite{wang2024non}, we substituted -1-valent Cl anion at the O site and Ni at the inner-layer Co site to achieve the same 3d$^{7.33}$ electronic configuration for both Co and Ni. The doped crystal structure La$_4$Co$_2$NiO$_8$Cl$_2$ (LCO-NiCl) is presented in Fig.~\ref{fig:structure}(c).

In this paper, we systematically investigate the electronic structure and correlation effect in the doped trilayer cobaltate LCO-NiCl using DFT+DMFT calculations, with undoped LNO-4310 as a reference. Our results reveal that LCO-NiCl exhibits electronic structure and correlation strength similar to those of superconducting LNO-4310. We find pronounced layer-dependent correlation, where the outer-layer Co $d_{z^2}$ orbitals show strong effective mass enhancement while the inner-layer Ni remains weakly correlated. The system also displays orbital-selective behavior characteristic of Hund's metal physics and flat bands near the Fermi level at the M point. Notably, the strong correlation character that resides in the outer-layer Ni sites of LNO-4310 is found in the outer-layer Co of LCO-NiCl upon doping. These findings establish LCO-NiCl as a compelling candidate for hosting high-temperature superconductivity in cobalt-based layered compounds and provide a theoretical foundation for future experimental exploration.

%%%%%%%%%%%%%%%%%%%%%%%%%%%%%%%%%%%%%%%%%%%%%%%%%%%%%%%%%%%%%%%%%%%%%%%%%%%%%%%%%%
%%--------------------------------- Section 2 ----------------------------------%%
%%%%%%%%%%%%%%%%%%%%%%%%%%%%%%%%%%%%%%%%%%%%%%%%%%%%%%%%%%%%%%%%%%%%%%%%%%%%%%%%%%
\section{METHODS}
In this study, we performed DFT+DMFT calculations with full charge-self-consistency using the eDMFT package~\cite{haule2010dynamical,
haule2015free} based on the WIEN2K code ~\cite{blaha2020wien2k}. Before performing DFT+DMFT, we first used the Vienna $ab$ $initio$ simulation package (VASP) for structural optimization and electronic structure calculations~\cite{kresse1994norm,
kresse1996efficiency}. The exchange-correlation energy was described by the generalized gradient approximation (GGA) with the Perdew-Burke-Ernzerhof (PBE) functional~\cite{perdew1996generalized}. For the VASP structural relaxations, we optimized the doped configuration LCO-NiCl based on the high-pressure tetragonal $I4/mmm$ phase of LNO-4310 at 44.3GPa~\cite{graser2009near,kemper2010sensitivity,li2023structural}.

In the DFT part, the muffin-tin radii (RMT) were set to 2.34, 1.82, 1.83, 1.57, and 2.13 bohr for La, Ni, Co, O, and Cl, respectively.  The maximum modulus for the reciprocal vectors $K_{max}$ was chosen such that $R_{MT} \times K_{max}=7.0$. Brillouin zone integrations were performed using a $k$ mesh of $12\times12\times12$. And in the DMFT part, we considered only the two e$_g$ orbitals ($d_{z^2}$ and $d_{x^2-y^2}$) of Ni and Co as correlated, since the $t_{2g}$ orbitals are fully occupied. The Coulomb interaction parameters were set to $U = 5.0$~eV and $J_H = 1.0$~eV for both Ni and Co 3$d$ electrons, following previous DFT+DMFT studies on nickelate superconductors~\cite{gu2025effective,anisimov1991band,cao2024flat}. We used the continuous-time quantum Monte Carlo (CT-QMC) solver to solve the impurity problem at a temperature of 290 K ($\beta = 40$~eV$^{-1}$), with $3\times10^7$ Monte Carlo steps to ensure convergence~\cite{gull2011continuous}. The exact double-counting scheme was used for the self-energy correction~\cite{haule2015exact}.

Since the LCO-NiCl contains multiple inequivalent transition metal sites, we performed site-resolved DFT+DMFT calculations with separate impurity problems for each inequivalent site. The LCO-NiCl configuration contains two inequivalent sites, namely the inner-layer Ni and the outer-layer Co. The self-energies on the real frequency axis were obtained by analytical continuation using the maximum entropy method~\cite{jarrell1996bayesian}. The mass enhancement was evaluated as ${m^*/m}=1 / Z=1-\left.\left(\partial \operatorname{Re} \sum(\omega) / \partial \omega\right)\right|_{\omega=0}$~\cite{gull2011continuous}. All calculations were performed in the paramagnetic state and spin-orbit coupling effects were neglected.

%%%%%%%%%%%%%%%%%%%%%%%%%%%%%%%%%%%%%%%%%%%%%%%%%%%%%%%%%%%%%%%%%%%%%%%%%%%%%%%%%%
%%--------------------------------- Section 3 ----------------------------------%%
%%%%%%%%%%%%%%%%%%%%%%%%%%%%%%%%%%%%%%%%%%%%%%%%%%%%%%%%%%%%%%%%%%%%%%%%%%%%%%%%%%
\section{RESULTS}
% {\it Results.}
% % \boxed{\text{Delta-Electron-Hole Pairing}}\\
% % \boxed{\text{Non-Bound Delta-Offset Pairing}}\\
% % \boxed{\text{Inter-orbital shiftting}}\\
% % \boxed{\text{doublon-holon excitations}}\\
Similar to the LNO-4310, the LCO-NiCl based on LCO-4310 also adopts the tetragonal $I4/mmm$ phase under high pressure of 44.3 GPa. The optimized lattice parameters obtained from structural relaxation using VASP are a = b = 3.5611 $\rm{\AA}$, c = 26.0039 $\rm{\AA}$ for LNO-4310; a = b = 3.6116 $\rm{\AA}$, c = 27.7503 $\rm{\AA}$ for the LCO-NiCl. The detailed crystal structures, including the NiO$_6$ and CoO$_6$ octahedra, are shown in Fig.~\ref{fig:structure}.

\subsection{Self-energy analysis}
\begin{figure*}[htb]
  \centering
  \includegraphics[width=\textwidth]{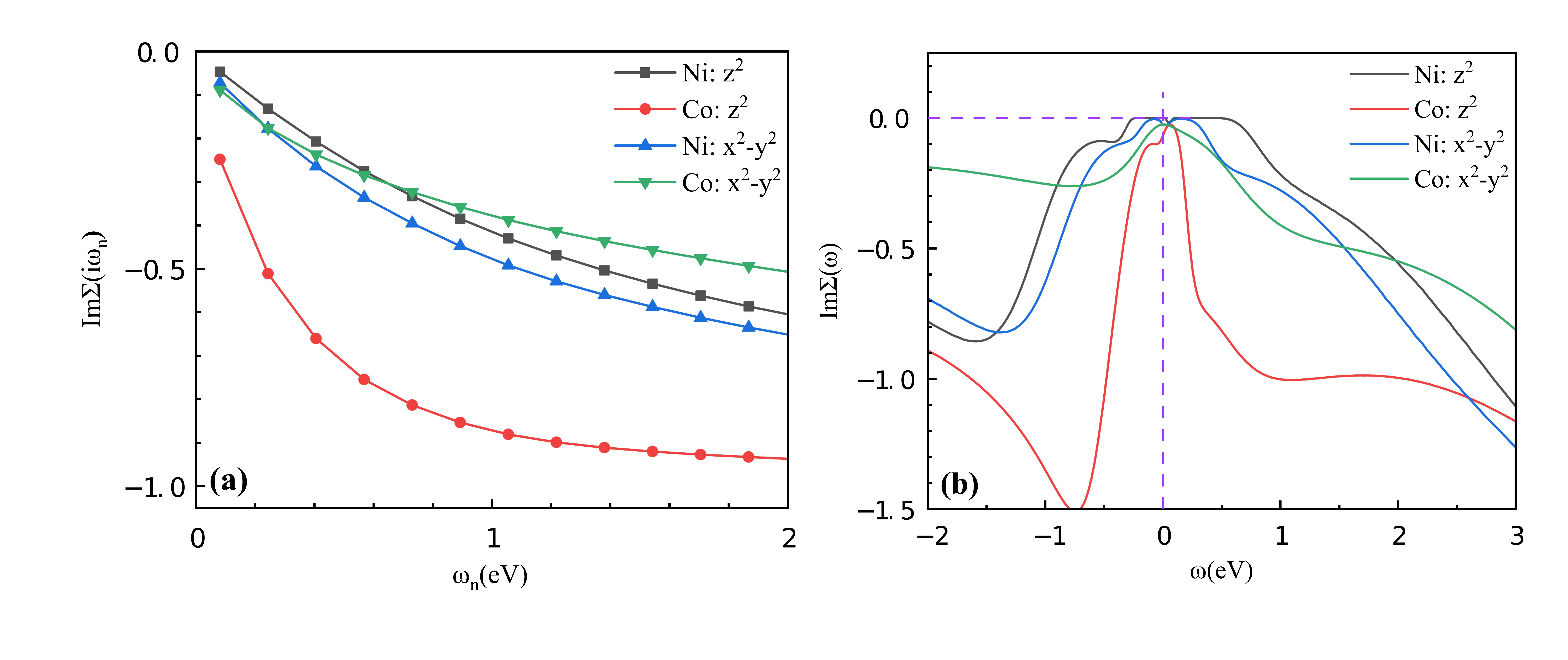}
  \caption{Imaginary parts of the self-energy at 290 K for LCO-NiCl. (a) Matsubara-frequency self-energy $\mathrm{Im\Sigma(i\omega_n)}$. (b) Real-frequency self-energy $\mathrm{Im\Sigma(\omega)}$.}
  \label{fig:self-energy}
\end{figure*}
 The self-energy of LCO-NiCl shown in Fig.~\ref{fig:self-energy} reveals clear orbital selectivity and layer-dependent correlation from both Matsubara-frequency and real-frequency self-energies. For the outer-layer Co, both $e_g$ orbitals exhibit finite $\mathrm{Im}\Sigma(i\omega_n)$ magnitudes at low frequencies and finite $\mathrm{Im}\Sigma(\omega)$ at $\omega = 0$, indicating non-Fermi liquid behavior, with the $d_{z^2}$ orbital exhibiting the largest magnitude among all orbitals and thus the strongest correlation. In contrast, the $\mathrm{Im\Sigma(i\omega_n)}$ magnitudes of both Ni $e_g$ orbitals are small and exhibit linear behavior at low frequencies, and their $\mathrm{Im\Sigma(\omega)}$ are close to zero near $\mathrm{\omega = 0}$, characteristic of Fermi liquid behavior. These observations demonstrate that in LCO-NiCl, strong correlation is confined to the outer-layer Co $d_{z^2}$ orbital, reflecting both orbital selectivity and layer-dependent electronic correlation.

This layer-dependent and orbital-selective correlation behavior closely resembles that of LNO-4310. In LNO-4310 under high pressure, the outer-layer Ni $d_{z^2}$ orbitals exhibit the strongest correlation with large $\mathrm{Im\Sigma(i\omega_n)}$ magnitudes, while the inner-layer Ni displays Fermi liquid characteristics with $\mathrm{Im\Sigma(\omega)}$ close to zero near the Fermi level and weaker correlation~\cite{wang2024non}. In both materials, the correlation character is confined to the outer layers and exhibits strong orbital selectivity. Specifically, the outer-layer $d_{z^2}$ orbitals are significantly more correlated than their $d_{x^2-y^2}$ counterparts and the substituted Ni in the inner layer of LCO-NiCl becomes weakly correlated, which further confirming our doping strategy successfully reproduces a correlation structure analogous to that of the nickelate superconductor.

\subsection{Momentum-resolved spectral functions and density of states}
\begin{figure*}[htb]
  \centering
  \includegraphics[width=\textwidth]{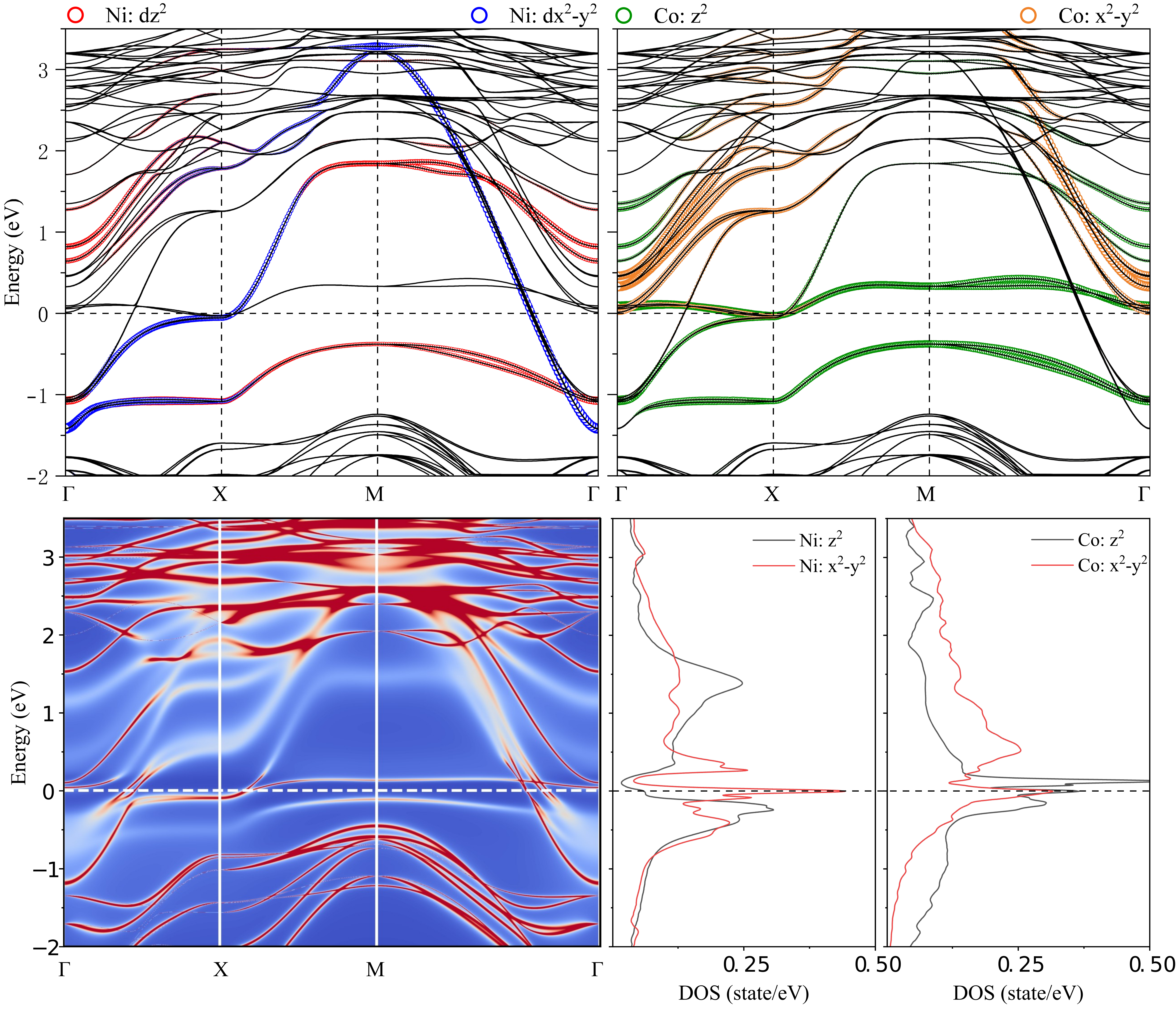}
  \caption{ DFT band structures are shown in the upper panels, where the orbital weights are represented by the size of the colored circles. The momentum-resolved spectral functions $\mathrm{A(\mathbf{k},\omega)}$ and density of states obtained by DFT+DMFT at 290 K are displayed in the lower panels. The white and black dashed lines at 0 eV denote the Fermi level.}
  \label{fig:AKW}
\end{figure*}
For the DFT+DMFT momentum-resolved spectral function of LCO-NiCl shown in Fig.~\ref{fig:AKW}, clear band renormalization effects are observed near the M point, where two flat bands appear on both sides of the Fermi level. Compared to the DFT bands, these DMFT spectral features exhibit much weaker dispersion, appearing significantly flatter, indicating the emergence of strong correlation. The energy bands of the Co and Ni $e_g$ orbitals are mainly distributed in the range of $-$1.5 to 3 eV, which are broadened and narrowed by correlation-induced renormalization. These flat bands near the Fermi level are predominantly contributed by the outer-layer Co dz² orbitals, analogous to LNO-4310 where similar flat bands originate from the outer-layer Ni $d_{z^2}$ orbitals~\cite{wang2024non}.

The corresponding DOS reveals distinct orbital characteristics. For Ni, the $d_{x^2-y^2}$ orbital has substantial weight near the Fermi level with a relatively high DOS value, indicating that this orbital plays a dominant role in the metallic behavior, while the Ni $d_{z^2}$ orbital shows a valley at $E_F$ with its main peak located at approximately -0.25 eV. For Co, the $d_{x^2-y^2}$ orbital exhibits a sharp peak at the Fermi level, indicating significant contribution to low-energy excitations. The Co $d_{z^2}$ orbital shows a small peak near $E_F$ with a value slightly larger than that of Co $d_{x^2-y^2}$, while its main peak is located approximately 0.1 eV above the Fermi level. These features demonstrate that in LCO-NiCl, the metallic conductivity is primarily governed by the Ni $d_{x^2-y^2}$ orbital and both Co $e_g$ orbitals, whereas the Ni $d_{z^2}$ orbital shows a valley at the Fermi level, reflecting its orbital-selective character.

Similar band renormalization and orbital selectivity are also observed in LNO-4310, where flat bands near the M point at the Fermi level are clearly visible and mainly originate from the Ni $d_{z^2}$ orbitals, exhibiting strong band renormalization due to correlation effect~\cite{wang2024non}. Both systems show pronounced band narrowing and renormalization compared to their DFT band structures, demonstrates that LCO-NiCl captures the analogous electronic characteristics of LNO-4310. These flat bands have been linked to unconventional superconducting pairing mechanisms in nickelates~\cite{shilenko2023correlated,zhang2024s}, indicating that similar physics may emerge in this cobalt-based analog. The presence of these features further supports the electronic similarity between the two materials and highlights the potential of LCO-NiCl as a cobalt-based counterpart to the nickelate superconductor.

\subsection{Effective mass enhancement and local multiplet}

To further characterize the orbital selectivity in LCO-NiCl, we analyze the effective mass enhancement obtained from our DFT+DMFT calculations, together with the orbital occupation numbers listed in Table~\ref{tab1}. These quantities reveal pronounced orbital selectivity in LCO-NiCl. The Co $d_{z^2}$ orbital exhibits a large effective mass enhancement, while its $d_{x^2-y^2}$ orbital shows a much smaller value, indicating strong orbital selectivity. The Ni orbitals in LCO-NiCl have effective mass enhancement values substantially reduced, suggesting that the substituted Ni in the inner layer becomes weakly correlated. In addition, these observations also reveal clear layer-dependent correlation, with strong correlation concentrated in the outer-layer Co and weak correlation in the inner-layer Ni. The occupation numbers provide additional evidence for these characteristics, with the two $e_g$ orbitals in both Co and Ni showing distinct occupation features that correlate with their different effective mass enhancement. The same orbital selectivity and layer-dependent correlation are also present in LNO-4310, where the outer-layer Ni orbitals exhibit larger effective mass enhancement than their inner-layer counterparts, with the $d_{z^2}$ orbital more strongly correlated than the $d_{x^2-y^2}$ orbital~\cite{wang2024non}. These similarities confirm that LCO-NiCl exhibits correlation physics similar to that of its nickelate counterpart.

A different perspective comes from the local spin multiplet weights of LCO-NiCl and LNO-4310, which are listed in Table~\ref{tab:multiplet_weights}. Previous theoretical studies on Ruddlesden-Popper nickelates have shown that strong local spin fluctuations are favorable for superconductivity~\cite{ouyang2025phase}. In the known superconductor LNO-4310, both outer- and inner-layer Ni exhibit a mixture of high-spin states ($N_\Gamma=2$, $S_z=1$) with low-spin ($N_\Gamma=1$, $S_z=1/2$) and three-electron ($N_\Gamma=3$, $S_z=1/2$) configurations, indicative of strong local spin fluctuations. This feature is also evident in LCO-NiCl, where the outer-layer Co and inner-layer Ni show high-spin states with weights of 27.7\% and 28.4\%, respectively, with multiple spin states coexisting. This similarity indicates that LCO-NiCl possesses a spin fluctuation characteristic similar to that of LNO-4310, which is favorable for unconventional superconductivity.

\begin{table}[t]
  \centering
  \renewcommand\arraystretch{1.5}
  \caption{Local orbital occupation numbers $N_d$ and effective mass enhancement $1/Z$ of the Ni and Co $d_{z^2}$ and $d_{x^2-y^2}$ orbitals for LNO-4310 and LCO-NiCl at 290 K. The LNO-4310 data are taken from Ref.~\cite{wang2024non}.}
  \label{tab1}
  \begin{tabular*}{8.3cm}{@{\extracolsep{\fill}} lcccccc}
    \hline\hline
    \multirow{2}{*}{} & 
    \multirow{2}{*}{Orbitals} &
    \multicolumn{2}{c}{LNO-4310} & 
    \multicolumn{2}{c}{LCO-NiCl} & \\
    \cmidrule(lr){3-4} \cmidrule(lr){5-6} 
    & & inner & outer & Ni & Co  \\
    \hline
    $N_d$  &$d_{z^2}$     & 1.060 & 1.088 & 1.108 & 1.017 \\
    $ $    &$d_{x^2-y^2}$ & 1.017 & 1.060 & 1.093 & 0.729 \\
    $ $    &total         & 2.077 & 2.148 & 2.201 & 1.746 \\
    \hline
    $1/Z$  &$d_{z^2}$     & 2.018 & 2.578 & 1.469 & 2.269 \\
    $ $    &$d_{x^2-y^2}$ & 1.986 & 2.296 & 1.540 & 1.456 \\
    \hline\hline
  \end{tabular*}
\end{table}
\begin{table}[t]
  \centering
  \renewcommand\arraystretch{1.5}
  \caption{Weights ($\%$) of the local multiplets for the inner- and outer-layer $e_g$ orbitals for LCO-NiCl and LNO-4310 at 290 K. The results for LNO-4310 are taken from Ref.~\cite{wang2024non}. The good quantum numbers $N_\Gamma$ and $S_z$ denote the total occupancy and total spin of the Ni and Co $e_g$ orbitals, which are used to label different local spin states. }
  \label{tab:multiplet_weights}
  \begin{tabular*}{8.3cm}{@{\extracolsep{\fill}} lccccccc}
    \hline\hline
    $N_\Gamma$ & 0 & 1 & 2 & 2 & 3 & 4 \\
    $S_z$ & 0 & 1/2 & 0 & 1 & 1/2 & 0 \\
    \hline
    Ni (LCO-NiCl) & 0.7 & 14.4 & 24.1 & 28.4 & 29.4 & 3.0 \\
    Co (LCO-NiCl) & 2.6 & 33.2 & 24.0 & 27.7 & 12.0 & 0.6 \\
    Ni-inner (LNO-4310) & 1.0 & 18.3 & 27.1 & 27.6 & 24.1 & 1.9 \\
    Ni-outer (LNO-4310) & 0.6 & 14.7 & 24.8 & 31.5 & 26.4 & 2.1 \\
    \hline\hline
  \end{tabular*}
\end{table}
\section{Discussion and conclusion}

The recent discovery of high-temperature superconductivity in the trilayer nickelate LNO-4310 under high pressure has naturally raised the question of whether similar physics can be realized in cobalt-based compounds. In this work, we aimed to engineer a cobaltate material whose electronic structure closely mimics that of LNO-4310, thereby establishing a potential platform for high-temperature superconductivity in Co-based systems.

In our initial attempt to achieve this goal, we pursued a direct electron-doping approach in La$_4$Co$_3$O$_{10}$ by substituting La with Th and O with Cl, aiming to increase the Co 3d electron count toward the 3d$^{7.33}$ configuration in LNO-4310. However, this approach failed, as the inner layer in the trilayer structure is intrinsically more hole-doped~\cite{wang2024non}. Consequently, electron doping via chemical substitution predominantly affected the outer layers, leaving the inner-layer Co largely unchanged with insufficient electron count and weak correlation.

To overcome this fundamental limitation, we adopted a site-selective substitution strategy by replacing the inner-layer Co itself with Ni, leading to our designed compound LCO-NiCl. With this structure, we performed DFT+DMFT calculations to investigate its electronic structures and correlation effects. Our results reveal that LCO-NiCl exhibits four characteristics closely resembling those of superconducting LNO-4310: (i) layer-dependent correlation, where strong electronic correlation is predominantly confined to the outer-layer Co $d_{z^2}$ orbitals; (ii) flat bands near the Fermi level, a feature that has been linked to the emergence of superconductivity in nickelates, both experimentally~\cite{zhu2024superconductivity} and theoretically~\cite{shilenko2023correlated,zhang2024s}; (iii) pronounced orbital selectivity, manifested in the distinct effective mass enhancement and orbital occupancies of the $e_g$ orbitals; and (iv) strong local spin fluctuations, a characteristic favorable for unconventional superconductivity~\cite{ouyang2025phase}. It should be noted that we focus on the high-pressure phase here, as the ambient-pressure structure may host complex spin or charge orders that are not accurately described by DFT, analogous to the case of high-pressure nickelate superconductors.

In conclusion, we have theoretically demonstrated that LCO-NiCl exhibits electronic structures and correlation features similar to those of superconducting LNO-4310. By circumventing the limitations of direct electron doping through inner-layer Ni substitution, we have successfully tuned a cobalt-based analog that captures the essential correlation physics of the nickelate superconductor. Although the resulting compound retains one Ni in the inner layer and is thus not a fully cobalt-based material, the outer-layer Co sites host the strong correlation essential for superconductivity, suggesting that Co itself can play the active role in realizing unconventional superconductivity. These findings establish LCO-NiCl as a promising candidate for hosting high-temperature superconductivity in cobalt-based layered compounds and provide a clear theoretical foundation for future experimental exploration.

Building on these results, two possible directions deserve future investigation. In particular, the substitution strategy demonstrated here can be extended to other transition metals to further tune the electronic correlation. In addition, given that superconductivity in LNO-4310 emerges under high pressure, calculating the pressure response of LCO-NiCl would be valuable to examine how compression modifies its band widths, correlation strengths, and Fermi surface topology.

\begin{acknowledgments}
This work was supported by the National Key R$\&$D
Program of China (Grants No. 2024YFA1408601 and
No. 2024YFA1408602) and the National Natural Science
Foundation of China (Grant No. 12434009). J.X.W. was
also supported by the Outstanding Innovative Talents
Cultivation Funded Programs 2025 of Renmin University
 of China. Z.Y.L. was also supported by the Innovation
 Program for Quantum Science and Technology
(Grant No. 2021ZD0302402). Computational resources
were provided by the Physical Laboratory of High Performance
 Computing in Renmin University of China.

\end{acknowledgments}

%\nocite{*}
\bibliography{lco}% Produces the bibliography via BibTeX.

@article{chu1987superconductivity,
  title={Superconductivity at 52.5 {K} in the lanthanum-barium-copper-oxide system},
  author={Chu, CW and Hor, PH and Meng, RL and Gao, L and Huang, ZJ},
  journal={Science},
  volume={235},
  number={4788},
  pages={567--569},
  year={1987},
  publisher={American Association for the Advancement of Science}
}

@article{schilling1993superconductivity,
  title={Superconductivity above 130 {K} in the hg--ba--ca--cu--o system},
  author={Schilling, Andreas and Cantoni, M and Guo, JD and Ott, HR},
  journal={Nature},
  volume={363},
  number={6424},
  pages={56--58},
  year={1993},
  publisher={Nature Publishing Group UK London}
}

@article{putilin1993superconductivity,
  title={Superconductivity at 94 {K} in {$\rm{Hg}\rm{Ba}_2\rm{Cu0}_{4+ \delta}$}},
  author={Putilin, SN and Antipov, EV and Chmaissem, Ol and Marezio, M},
  journal={Nature},
  volume={362},
  number={6417},
  pages={226--228},
  year={1993},
  publisher={Nature Publishing Group UK London}
}

@article{park1995structures,
  title={Structures of High-Temperature Cuprate Superconductors},
  author={Park, Chan and Snyder, Robert L},
  journal={Journal of the American Ceramic Society},
  volume={78},
  number={12},
  pages={3171--3194},
  year={1995},
  publisher={Wiley Online Library}
}

@article{kamihara2008iron,
  title={Iron-Based Layered Superconductor {La[O$_{1-x}$F$_x$]FeAs} ({$x$} = 0.05--0.12) with {$T_{\rm c}$} = 26 {K}},
  author={Kamihara, Yoichi and Watanabe, Takumi and Hirano, Masahiro and Hosono, Hideo},
  journal={Journal of the American Chemical Society},
  volume={130},
  number={11},
  pages={3296--3297},
  year={2008},
  publisher={ACS Publications}
}

@article{rotter2008superconductivity,
  title={Superconductivity at 38 {K} in the iron arsenide {(Ba$_{1-x}$K$_x$)Fe$_2$As$_2$}},
  author={Rotter, Marianne and Tegel, Marcus and Johrendt, Dirk},
  journal={Physical review letters},
  volume={101},
  number={10},
  pages={107006},
  year={2008},
  publisher={APS}
}

@article{johnston2010puzzle,
  title={The puzzle of high temperature superconductivity in layered iron pnictides and chalcogenides},
  author={Johnston, David C},
  journal={Advances in Physics},
  volume={59},
  number={6},
  pages={803--1061},
  year={2010},
  publisher={Taylor \& Francis}
}

@article{botana2020similarities,
  title={Similarities and differences between LaNiO 2 and CaCuO 2 and implications for superconductivity},
  author={Botana, Antia S and Norman, Michael R},
  journal={Physical Review X},
  volume={10},
  number={1},
  pages={011024},
  year={2020},
  publisher={APS}
}

@article{poltavets2009electronic,
  title={Electronic Properties, Band Structure, and Fermi Surface Instabilities of Ni 1+/Ni 2+ Nickelate La 3 Ni 2 O 6, Isoelectronic with Superconducting Cuprates},
  author={Poltavets, Viktor V and Greenblatt, Martha and Fecher, Gerhard H and Felser, Claudia},
  journal={Physical review letters},
  volume={102},
  number={4},
  pages={046405},
  year={2009},
  publisher={APS}
}

@article{greenblatt2010bulk,
  title={Bulk Magnetic Order in a Two Dimensional Ni\^{} 1+/Ni\^{} 2+(d\^{} 9/d\^{} 8) Nickelate, Isoelectronic with Superconducting Cuprates},
  author={Greenblatt, Martha},
  journal={Materials by Design: Understanding and Controlling the Electronic Properties of Novel Correlated Electron Systems},
  pages={9},
  year={2010}
}

@article{sakakibara2020model,
  title={Model construction and a possibility of cupratelike pairing in a new d 9 nickelate superconductor (Nd, Sr) NiO 2},
  author={Sakakibara, Hirofumi and Usui, Hidetomo and Suzuki, Katsuhiro and Kotani, Takao and Aoki, Hideo and Kuroki, Kazuhiko},
  journal={Physical Review Letters},
  volume={125},
  number={7},
  pages={077003},
  year={2020},
  publisher={APS}
}

@article{wang2025prediction,
  title={Prediction of several Co-based La $ \_3 $ Ni $ \_2 $ O $ \_7 $-like superconducting materials},
  author={Wang, Jing-Xuan and Tian, Yi-Heng and She, Jian-Hong and He, Rong-Qiang and Lu, Zhong-Yi},
  journal={arXiv preprint arXiv:2509.09664},
  year={2025}
}

@article{sun2023signatures,
  title={Signatures of superconductivity near 80 {K} in a nickelate under high pressure},
  author={Sun, Hualei and Huo, Mengwu and Hu, Xunwu and Li, Jingyuan and Liu, Zengjia and Han, Yifeng and Tang, Lingyun and Mao, Zhongquan and Yang, Pengtao and Wang, Bosen and others},
  journal={Nature},
  volume={621},
  number={7979},
  pages={493--498},
  year={2023},
  publisher={Nature Publishing Group UK London}
}

@article{wang2024pressure,
  title={Pressure-induced superconductivity in polycrystalline {La$_3$Ni$_2$O$_{7-\delta}$}},
  author={Wang, Gang and Wang, NN and Shen, XL and Hou, Jun and Ma, Liang and Shi, LF and Ren, ZA and Gu, YD and Ma, HM and Yang, PT and others},
  journal={Physical Review X},
  volume={14},
  number={1},
  pages={011040},
  year={2024},
  publisher={APS}
}

@article{zhou2025investigations,
  title={Investigations of key issues on the reproducibility of high-Tc superconductivity emerging from compressed {La$_3$Ni$_2$O$_7$}},
  author={Zhou, Yazhou and Guo, Jing and Cai, Shu and Sun, Hualei and Li, Chengyu and Zhao, Jinyu and Wang, Pengyu and Han, Jinyu and Chen, Xintian and Chen, Yongjin and others},
  journal={Matter and Radiation at Extremes},
  volume={10},
  number={2},
  year={2025},
  publisher={AIP Publishing}
}

@article{luo2023bilayer,
  title={Bilayer two-orbital model of {La$_3$Ni$_2$O$_7$} under pressure},
  author={Luo, Zhihui and Hu, Xunwu and Wang, Meng and W{\'u}, W{\'e}i and Yao, Dao-Xin},
  journal={Physical review letters},
  volume={131},
  number={12},
  pages={126001},
  year={2023},
  publisher={APS}
}

@article{zhu2024superconductivity,
  title={Superconductivity in pressurized trilayer {La$_4$Ni$_3$O$_{10-\delta}$} single crystals},
  author={Zhu, Yinghao and Peng, Di and Zhang, Enkang and Pan, Bingying and Chen, Xu and Chen, Lixing and Ren, Huifen and Liu, Feiyang and Hao, Yiqing and Li, Nana and others},
  journal={Nature},
  volume={631},
  number={8021},
  pages={531--536},
  year={2024},
  publisher={Nature Publishing Group UK London}
}

@article{li2024signature,
  title={Signature of superconductivity in pressurized {La$_4$Ni$_3$O$_{10}$}},
  author={Li, Qing and Zhang, Ying-Jie and Xiang, Zhe-Ning and Zhang, Yuhang and Zhu, Xiyu and Wen, Hai-Hu},
  journal={Chinese Physics Letters},
  volume={41},
  number={1},
  pages={017401},
  year={2024},
  publisher={Chinese Physical Society and IOP Publishing Ltd}
}

@article{shilenko2023correlated,
  title={Correlated electronic structure, orbital-selective behavior, and magnetic correlations in double-layer {La$_3$Ni$_2$O$_7$} under pressure},
  author={Shilenko, DA and Leonov, IV},
  journal={Physical Review B},
  volume={108},
  number={12},
  pages={125105},
  year={2023},
  publisher={APS}
}

@article{cao2024flat,
  title={Flat bands promoted by Hund's rule coupling in the candidate double-layer high-temperature superconductor {La$_3$Ni$_2$O$_7$} under high pressure},
  author={Cao, Yingying and Yang, Yi-feng},
  journal={Physical Review B},
  volume={109},
  number={8},
  pages={L081105},
  year={2024},
  publisher={APS}
}

@article{leonov2024electronic,
  title={Electronic structure and magnetic correlations in the trilayer nickelate superconductor {La$_4$Ni$_3$O$_{10}$} under pressure},
  author={Leonov, IV},
  journal={Physical Review B},
  volume={109},
  number={23},
  pages={235123},
  year={2024},
  publisher={APS}
}

@article{sakakibara2024theoretical,
  title={Theoretical analysis on the possibility of superconductivity in the trilayer Ruddlesden-Popper nickelate {La$_4$Ni$_3$O$_{10}$} under pressure and its experimental examination: Comparison with {La$_3$Ni$_2$O$_7$}},
  author={Sakakibara, Hirofumi and Ochi, Masayuki and Nagata, Hibiki and Ueki, Yuta and Sakurai, Hiroya and Matsumoto, Ryo and Terashima, Kensei and Hirose, Keisuke and Ohta, Hiroto and Kato, Masaki and others},
  journal={Physical Review B},
  volume={109},
  number={14},
  pages={144511},
  year={2024},
  publisher={APS}
}

@article{li2023structural,
  title={Structural transition and electronic band structures in the compressed trilayer nickelate {La$_4$Ni$_3$O$_{10}$}},
  author={Li, Jingyuan and Chen, Cuiqun and Huang, Chaoxin and Han, Yifeng and Huo, Mengwu and Huang, Xing and Ma, Peiyue and Qiu, Zhengyang and Chen, Junfeng and Chen, Lan and others},
  journal={arXiv e-prints},
  pages={arXiv--2311},
  year={2023}
}

@article{sakakibara2024possible,
  title={Possible high T c superconductivity in {La$_3$Ni$_2$O$_7$} under high pressure through manifestation of a nearly half-filled bilayer Hubbard model},
  author={Sakakibara, Hirofumi and Kitamine, Naoya and Ochi, Masayuki and Kuroki, Kazuhiko},
  journal={Physical Review Letters},
  volume={132},
  number={10},
  pages={106002},
  year={2024},
  publisher={APS}
}

@article{wang2024non,
  title={Non-Fermi liquid and Hund correlation in {La$_4$Ni$_3$O$_{10}$} under high pressure},
  author={Wang, Jing-Xuan and Ouyang, Zhenfeng and He, Rong-Qiang and Lu, Zhong-Yi},
  journal={Physical Review B},
  volume={109},
  number={16},
  pages={165140},
  year={2024},
  publisher={APS}
}

@article{zhang2025superconductivity,
  title={Superconductivity in trilayer nickelate {La$_4$Ni$_3$O$_{10}$} under pressure},
  author={Zhang, Mingxin and Pei, Cuiying and Peng, Di and Du, Xian and Hu, Weixiong and Cao, Yantao and Wang, Qi and Wu, Juefei and Li, Yidian and Liu, Huanyu and others},
  journal={Physical Review X},
  volume={15},
  number={2},
  pages={021005},
  year={2025},
  publisher={APS}
}

@article{ouyang2025phase,
  title={Phase diagrams and two key factors to superconductivity of Ruddlesden-Popper nickelates},
  author={Ouyang, Zhenfeng and He, Rong-Qiang and Lu, Zhong-Yi},
  journal={Physical Review B},
  volume={112},
  number={4},
  pages={045127},
  year={2025},
  publisher={APS}
}

@article{hu2015predicting,
  title={Predicting unconventional high-temperature superconductors in trigonal bipyramidal coordinations},
  author={Hu, Jiangping and Le, Congcong and Wu, Xianxin},
  journal={Physical Review X},
  volume={5},
  number={4},
  pages={041012},
  year={2015},
  publisher={APS}
}

@article{armitage2010progress,
  title={Progress and perspectives on electron-doped cuprates},
  author={Armitage, NP and Fournier, P and Greene, RL},
  journal={Reviews of Modern Physics},
  volume={82},
  number={3},
  pages={2421--2487},
  year={2010},
  publisher={APS}
}

@article{graser2009near,
  title={Near-degeneracy of several pairing channels in multiorbital models for the Fe pnictides},
  author={Graser, Siegfried and Maier, TA and Hirschfeld, PJ and Scalapino, DJ},
  journal={New Journal of Physics},
  volume={11},
  number={2},
  pages={025016},
  year={2009}
}

@article{kemper2010sensitivity,
  title={Sensitivity of the superconducting state and magnetic susceptibility to key aspects of electronic structure in ferropnictides},
  author={Kemper, Alexander F and Maier, Thomas A and Graser, Siegfried and Cheng, Hai-Ping and Hirschfeld, PJ and Scalapino, DJ},
  journal={New Journal of Physics},
  volume={12},
  number={7},
  pages={073030},
  year={2010}
}

@article{haule2010dynamical,
  title={Dynamical mean-field theory within the full-potential methods: Electronic structure of {CeIrIn$_5$, CeCoIn$_5$, and CeRhIn$_5$}},
  author={Haule, Kristjan and Yee, Chuck-Hou and Kim, Kyoo},
  journal={Physical Review B—Condensed Matter and Materials Physics},
  volume={81},
  number={19},
  pages={195107},
  year={2010},
  publisher={APS}
}

@article{haule2015free,
  title={Free energy from stationary implementation of the {DFT+ DMFT} functional},
  author={Haule, Kristjan and Birol, Turan},
  journal={Physical review letters},
  volume={115},
  number={25},
  pages={256402},
  year={2015},
  publisher={APS}
}

@article{blaha2020wien2k,
  title={{WIEN2k}: An {APW+ lo} program for calculating the properties of solids},
  author={Blaha, Peter and Schwarz, Karlheinz and Tran, Fabien and Laskowski, Robert and Madsen, Georg KH and Marks, Laurence D},
  journal={The Journal of chemical physics},
  volume={152},
  number={7},
  year={2020},
  publisher={AIP Publishing}
}

@article{kresse1994norm,
  title={Norm-conserving and ultrasoft pseudopotentials for first-row and transition elements},
  author={Kresse, Georg and Hafner, Jurgen},
  journal={Journal of Physics: Condensed Matter},
  volume={6},
  number={40},
  pages={8245--8257},
  year={1994}
}

@article{kresse1996efficiency,
  title={Efficiency of ab-initio total energy calculations for metals and semiconductors using a plane-wave basis set},
  author={Kresse, Georg and Furthm{\"u}ller, J{\"u}rgen},
  journal={Computational materials science},
  volume={6},
  number={1},
  pages={15--50},
  year={1996},
  publisher={Elsevier}
}

@article{perdew1996generalized,
  title={Generalized gradient approximation made simple},
  author={Perdew, John P and Burke, Kieron and Ernzerhof, Matthias},
  journal={Physical review letters},
  volume={77},
  number={18},
  pages={3865},
  year={1996},
  publisher={APS}
}

@article{gu2025effective,
  title={Effective model and pairing tendency in the bilayer Ni-based superconductor {La$_3$Ni$_2$O$_7$}},
  author={Gu, Yuhao and Le, Congcong and Yang, Zhesen and Wu, Xianxin and Hu, Jiangping},
  journal={Physical Review B},
  volume={111},
  number={17},
  pages={174506},
  year={2025},
  publisher={APS}
}

@article{anisimov1991band,
  title={Band theory and Mott insulators: Hubbard {U} instead of Stoner {I}},
  author={Anisimov, Vladimir I and Zaanen, Jan and Andersen, Ole K},
  journal={Physical Review B},
  volume={44},
  number={3},
  pages={943},
  year={1991},
  publisher={APS}
}

@article{gull2011continuous,
  title={{Continuous-time Monte Carlo} methods for quantum impurity models},
  author={Gull, Emanuel and Millis, Andrew J and Lichtenstein, Alexander I and Rubtsov, Alexey N and Troyer, Matthias and Werner, Philipp},
  journal={Reviews of modern physics},
  volume={83},
  number={2},
  pages={349--404},
  year={2011},
  publisher={APS}
}

@article{jarrell1996bayesian,
  title={Bayesian inference and the analytic continuation of imaginary-time quantum {Monte Carlo} data},
  author={Jarrell, Mark and Gubernatis, James E},
  journal={Physics Reports},
  volume={269},
  number={3},
  pages={133--195},
  year={1996},
  publisher={Elsevier}
}

@article{haule2015exact,
  title={Exact double counting in combining the dynamical mean field theory and the density functional theory},
  author={Haule, Kristjan},
  journal={Physical review letters},
  volume={115},
  number={19},
  pages={196403},
  year={2015},
  publisher={APS}
}

@article{zhang2024s,
  title={The $s^{\pm}$-Wave Superconductivity in the Pressurized La$_4$Ni$_3$O$_{10}$},
  author={Zhang, Ming and Sun, Hongyi and Liu, Yu-Bo and Liu, Qihang and Chen, Wei-Qiang and Yang, Fan},
  journal={arXiv preprint arXiv:2402.07902},
  year={2024}
}

\end{document}